\documentclass{aastex}
\newcommand{\rbh}{\boldsymbol {r}}

\newcommand{\taub}{\boldsymbol{\tau}}

\newcommand{\Ab}{\boldsymbol A}
\newcommand{\db}{\boldsymbol d}
\newcommand{\gb}{\boldsymbol g}

\newcommand{\kb}{\boldsymbol k}

\newcommand{\tb}{\boldsymbol t}

\newcommand{\rhob}{\boldsymbol{\rho}}
\newcommand{\Deltab}{\boldsymbol{\Delta}}
\newcommand{\Hbs}{\boldsymbol H^*}
\newcommand{\Hb}{\boldsymbol H}
\newcommand{\Mb}{\boldsymbol M}
\newcommand{\Rb}{\boldsymbol R}
\newcommand{\Sb}{\boldsymbol S}
\newcommand{\Jb}{\boldsymbol J}
\newcommand{\Xb}{\boldsymbol X}

\usepackage{amsbsy}
\shorttitle{Non-Gaussian Random Fields}
\shortauthors{Vio, Andreani, Tenorio, Wamsteker}

\begin{document}
\title{Numerical Simulation of Non-Gaussian Random Fields
\\ with Prescribed Marginal Distributions and Cross-Correlation Structure II: Multivariate Random Fields}
\author{Roberto Vio\altaffilmark{1}}
\affil{Chip Computers Consulting s.r.l., \\ Viale Don L.~Sturzo 82,
S.Liberale di Marcon, 30020 Venice, Italy}
\email{robertovio@tin.it}
\author{Paola Andreani\altaffilmark{2}}
\affil{Osservatorio Astronomico di Padova,\\ vicolo dell'Osservatorio 5,
35122 Padua, Italy}
\email{andreani@mpe.mpg.de}
\author{Luis Tenorio}
\affil{Department of Mathematical and Computer Sciences,\\ Colorado School
of Mines, Golden CO 80401, USA}
\email{ltenorio@mines.edu}
\author{Willem Wamsteker}
\affil{ESA-VILSPA, \\Apartado 50727, 28080 Madrid, Spain}
\email{willem.wamsteker@esa.int}
\altaffiltext{1}{ESA-VILSPA, Apartado 50727, 28080 Madrid,
Spain}
\altaffiltext{2}{Max-Planck Institut f\"ur Extraterrestrische Physik,
Postfach 1312
85741 Garching, Germany}

\begin{abstract}

We provide theoretical procedures and practical recipes to simulate non-Gaussian correlated,
homogeneous random fields with prescribed marginal distributions and cross-correlation
structure,
either in a $N$-dimensional Cartesian space or on the celestial
sphere. We illustrate our methods using far-infrared maps obtained with
the Infrared Space Observatory.
However, the methodology presented here can be used in other
astrophysical applications that require modeling correlated features in sky maps,
for example, the simulation of multifrequency sky maps where
backgrounds, sources and noise are correlated and can be modeled by random fields.

\end{abstract}

\section{Introduction}

Random fields are widely used in astrophysics to model realistic scenarios of
physical processes that depend on random components. For example,
they are widely used in cosmology to model different types of galactic and
extragalactic backgrounds
\citep[e.g.,][]{martinez}. The physical characteristics of the backgrounds are translated into
statistical characteristics of the field. The fields are usually required to be ergodic so
that information
about the model
can be extracted from a single realization of the field. In addition, they are assumed to be
isotropic (the correlation between
two points depends only on the distance that separates them) or homogeneous (the correlation
depends on the
difference of their position vectors) to reflect the geometry of the cosmological model.
Given the complexity of the cosmological models, characteristics of the fields are usually
studied
via Monte Carlo simulations where model predictions are compared with observations.

To simulate a Gaussian random field we only need to specify the mean and spectrum
(or correlation function)
but there are non-Gaussian models whose predictions also have to be tested against observations.
For example,
standard inflationary models predict Gaussian temperature fluctuations of the cosmic
microwave background but there are other cosmological models that predict non-Gaussian
fluctuations
\citep{avelino,peebles}. Homogeneous non-Gaussian random fields are more difficult to define
and simulate,
and since they are not uniquely determined by their first two moments, we often have to
accept only a partial second order description of the fields.

\citet{vio01} (hereafter {\it VAW}) presented numerical methods for the simulation of
homogeneous scalar
random fields $R(\tb)$
with prescribed one-dimensional distribution function (marginal distribution)
$ F_R(r)$ and correlation structure. These methods can be used, for example, to generate
cosmic microwave background maps with a given spectrum and a marginal
distribution that allows for asymmetry or kurtosis in the pixel temperatures.
In this
paper we generalize these methods to the case of pair-wise correlated random fields defined
on the same physical space. This will allow us, for example, to simulate backgrounds and
source fields that are not independent, as in star-forming regions
where the already formed stars are linked to the surrounding gaseous and dusty environment
from which
they originate.

We consider random fields defined in $N$-dimensional ``parameter spaces''
where the coordinates $\tb = (t_1,t_2,\ldots,t_N)$ may correspond
to spatial/angular coordinates (spatial random fields), time
(time processes), a mix of these two (spatio-temporal random fields), or to
even more general situations. We also consider the case of random fields defined on the
sphere,
that is, random fields that depend only on the direction in the sky.
If the multivariate random field is $\Rb(\tb) = \{ R_1(\tb)$, $R_2(\tb)$,$\ldots$,
$R_M(\tb) \}$, the goal is to generate a field with components $R_i$
having (possibly different) prescribed marginal distributions $F_{R_i}(r_i)$,
prescribed correlation functions, and prescribed pair-wise cross-correlations.

In theory, a complete description of a random field requires the definition
of all finite-dimensional joint distributions, but, unless
the field is Gaussian, this is a terribly difficult task.
In practical applications one only considers a second order description of
the field that specifies the marginals $F_{R_i}(r_i)$,
the means $\mu_{R_i}={\rm E}[\,R_i(\tb)\,]$ and the cross-covariance
functions
\begin{equation} \label{eq:covariance}
\xi_{R_i R_j}(\tb_1,\tb_2)= {\rm E} \left[\, (R_i(\tb_1)-\mu_{R_i})
~(R_j(\tb_2) -\mu_{R_j})\, \right],
\end{equation}
where ${\rm E}[\cdot]$ stands for expected value (ensemble average).

As mentioned before, it is often possible, even necessary,
to adopt some simplifying assumptions like isotropy or homogeneity of the
field $\Rb(\tb)$. In the multidimensional context, isotropy means that the cross-covariance
function depends
on the length $\tau=\Vert \taub \Vert$ of the vector $\taub= \tb_1-\tb_2$ but not on its
direction:
$\xi_{R_i R_j}(\tb_1,\tb_2)= \xi_{R_i R_j}(\Vert \tb_1-\tb_2 \Vert)$.
In this case the cross-correlation function (normalized covariance function)
depends only on $\tau$
\begin{equation} \label{eq:rhos}
\rhob_{\Rb}(\tau) = \left(
\begin{array}{ccc}
\rho_{R_1 R_1}(\tau) & &\rho_{R_1 R_M}(\tau) \\
\vdots & \ddots & \vdots \\
\rho_{R_M R_1}(\tau) & &\rho_{R_M R_M}(\tau)
\end{array} \right),
\end{equation}
where
\begin{equation} \label{eq:rho}
\rho_{R_i R_j}(\tau) = {\rm E} \left[ \frac{(R_i(\tb) - \mu_{R_i})
~(R_j(\tb + \taub)
- \mu_{R_j})}{\sigma_{R_i} \sigma_{R_j}} \right]
\end{equation}
and $\sigma_{R_i}^2$ is the variance corresponding to the
marginal $F_{R_i}(r_i)$. By definition, $\rho_{R_i R_i}(0) = 1$
for $i=1,2,\ldots,M$.

Although the isotropic case is of great interest in astronomical
applications, here we consider multidimensional homogeneous random fields in general, that is,
random fields whose covariance function depends on $\taub$.
In this case, the correlation function
$\rhob_{\Rb}(\taub)$ is defined as in equation
(\ref{eq:rhos}) but with $\taub$ instead of $\tau$.

The rest of the paper is organized as follows. In Section \ref{sec:notes} we describe a
general procedure for
generating non-Gaussian random fields by point-wise transformations of Gaussian ones.
In Section \ref{sec:practical} we provide practical recipes for the methods
outlined in Section \ref{sec:notes} and show how they can be used in
either $N$-dimensional Cartesian spaces or on the celestial sphere.
Section \ref{sec:examples} presents some examples. We show simulated maps of the
background and localized source field components of a far-IR sky map of
the Infrared Space Observatory (ISO). In the ISO map, the locations and intensities of the
sources are
correlated with the background field and only a multidimensional
simulation may reproduce such physical characteristic.

\section{Transforming Gaussian fields to non-Gaussian} \label{sec:notes}
In Section \ref{sec:grf} we discuss methods for generating homogeneous
Gaussian random fields with a prescribed correlation
structure. The procedures are straightforward, at least in principle, while,
in contrast, generating homogeneous non-Gaussian fields is not a trivial
task. Fortunately, a large family of such fields can be obtained via
pointwise transformations of homogeneous Gaussian fields. These
transformations are the basis of the methods used in {\it VAW} to simulate
non-Gaussian scalar random fields. In this paper we generalize them to the multivariate
case.

We propose the following procedure for simulating multidimensional homogeneous
non-Gaussian fields $\Rb(\tb)$ with prescribed cross-correlation
structure $\rhob_{\Rb}(\taub)$ and one-dimensional marginals $F_{R_i}(r)$:
(a) Generate a zero-mean, unit-variance, homogeneous multivariate Gaussian
field $\Xb(\tb)$ with a fixed cross-correlation structure
$\rhob_{\Xb}(\taub)$; (b)
transform $\Xb(\tb)$ to a non-Gaussian field $ \Rb(\tb)$ via
\begin{equation} \label{eq:map}
\Rb(\tb)= \gb [\, \Xb(\tb)\, ],
\end{equation}
where $\gb(\cdot) = \{ g_1(\cdot), g_2(\cdot), \ldots, g_M(\cdot) \}$.
The problem is to determine a Gaussian field $\Xb(\tb)$ and
functions $\{ g_i \}$ that transform $\rhob_{\Xb}(\taub)$
into $\rhob_{\Rb}(\taub)$.

\begin{figure}[h] 
\plotone{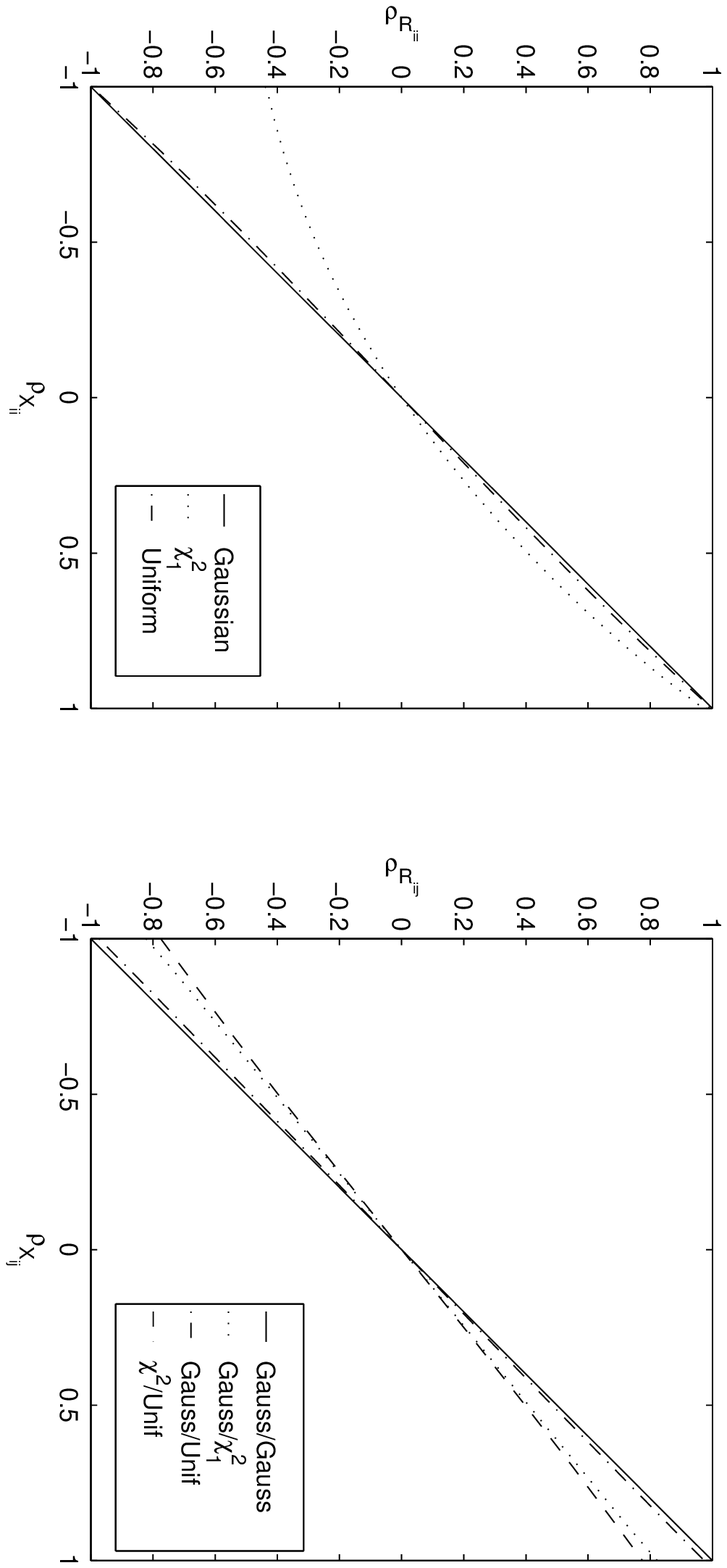}
\caption{Examples of associations between the
cross-correlation functions $\rhob_{\Rb}(\taub)$ and $\rhob_{\Xb}(\taub)$
given by equation (\ref{eq:change}) for some non-Gaussian random fields.}
\label{fig:rhos} 
\end{figure}

As discussed in {\it VAW}, given marginals $\{ F_{R_i}(r) \}$ with no atoms
(a concentration of a finite probability mass at a point), the integral
transform
\begin{equation} \label{eq:map1}
R_i(\tb)= g_i[\,X_i(\tb)\,]= F^{-1}_{R_i} \{ \,F_{X} [\, X_i(\tb)\, ]\, \};
\qquad i=1,2,\ldots, M,
\end{equation}
where $F_{X}$ denotes the standard Gaussian distribution function and
$F^{-1}_{R_i}$ is the inverse distribution function of $R_i(\tb)$, provides
a random field with marginals $\{ F_{R_i}(r) \}$. Since
$g_i(\cdot)=F^{-1}_{R_i} \{ F_X(\cdot) \}$ is a strictly monotonic function,
transformation (\ref{eq:map1}) is invertible and $\rho_{R_i R_j}(\taub)$ can
be calculated through
\begin{equation} \label{eq:change}
\rho_{R_i R_j}(\taub)= \frac{1}{\sigma_{R_i} \sigma_{R_j}}
\int_{-\infty}^{\infty}
\int_{-\infty}^{\infty} [g_i(x_1) - \mu_{R_i}] ~[g_j(x_2) - \mu_{R_j}]
~\phi(x_1, x_2; \rho_{X_i X_j}(\taub))~dx_1 ~dx_2,
\end{equation}
where, $x_1= x(\tb)$, $x_2= x(\tb+\taub)$, and
\begin{equation}
\phi(x_1, x_2;\, \rho_{X_i X_j}(\taub)\,)= \frac{1}{2 \pi\, [\,1
-\rho_{X_i X_j}^2(\taub)\,]^{1/2}} ~\exp \left(
- \frac{x_1^2 + x_2^2 - 2 \rho_{X_i X_j}(\taub)\, x_1 x_2}{2\, [\,1 -
\rho_{X_i X_j}^2(\taub)\,]} \right).
\end{equation}
For example, Figure \ref{fig:rhos} shows the relationship between
$\rho_{X_i X_j}$ and $\rho_{R_i R_j}$ for some standard marginal
distributions.

Equation (\ref{eq:change}) can be used to solve for the cross-correlations of
a Gaussian field that transform to those of the non-Gaussian field.
\citet{ogo96} show that $\rho_{R_i R_j}(\taub)$
is a monotonic increasing function of $\rho_{X_i X_j}(\taub)$
and that (\ref{eq:change}) does not present invertibility problems.
However, although via this inversion it is always possible to map a value of
$\rho_{X_i X_j}(\taub)$ to its corresponding value $\rho_{R_i R_j}(\taub)$,
there is no guarantee that the resulting $\rho_{X_i X_j}(\taub)$ is in fact
a non-negative definite function and hence a valid cross-correlation
function. This condition must be checked (see below).
There are also restrictions on the distributions $F_{R_i}$ that can
be obtained. For example, as already shown in {\it VAW}, the values of
$\rho_{R_i R_j}$ that can be inverted are those in the interval
$ \rho_{R_i R_j} \in [\,\rho^*_1, \rho^*_2\,]$,
where $\rho^*_1$ and $\rho^*_2$ correspond, respectively, to the values of
$-1$ and $1$ of $\rho_{X_i X_j}$ for a fixed function $\gb$. This is clearly seen
in Figure \ref{fig:rhos} where relationships between $ \rho_{R_i R_j}$ and
$ \rho_{X_i X_j} $ for some classic distribution functions are shown.

To conclude this section we note that the fields $R_i(\tb)$ can be forced to
have a chosen mean $\tilde \mu_i$ and variance $\tilde \sigma^2_i$ by applying the
transformation
\begin{equation}
\tilde R_i(\tb)= \tilde \mu_{R_i} + \frac{R_i(\tb) - \mu_{R_i}}{\sigma_{R_i}}
~~ \tilde \sigma_{R_i}.
\end{equation}
without modifying the cross-correlation structure.
\section{Practical recipes} \label{sec:practical}

To simulate a non-Gaussian field $\Rb(\tb)$ via equations
(\ref{eq:map1}) and (\ref{eq:change})
we need to (a) determine the appropriate cross-correlation functions
$\rho_{X_i X_j}(\taub)$ of the Gaussian fields $X_i(\tb)$ to be
transformed, (b) check that $\rhob_{\Xb}(\taub)$ is a non-negative
definite function, and (c) simulate the Gaussian fields $X_i(\tb)$.

\subsection{Solving the inverse problem} \label{sec:invprob}

In principle, $\rho_{X_i X_j}(\taub)$ can be obtained by inverting
equation (\ref{eq:change}) but this procedure is computationally expensive.
Instead, since the relationship between  $\rho_{X_i X_j}(\taub)$ and
$\rho_{R_i R_j}(\taub)$ is generally smooth, it is sufficient to choose
a set of values in the interval $[-1, ~ 1]$ for $\rho_{X_i X_j}(\taub)$ and
to solve for their corresponding values $\rho_{R_i R_j}(\taub)$ via the
numerical integration of equation (\ref{eq:change}).
The complete relationship between the two correlation functions may be obtained by
spline interpolation.
The interpolated function is then used to determine $\rho_{X_i X_j}(\taub)$ as a
function of $\taub$.

The function  $\rhob_{\Xb}(\taub)$ has to be non-negative definite to
correspond to a homogeneous field. This means that every
$\rho_{X_i X_j}(\taub)$ is a non-negative definite function and that the
matrix $\rhob_{\Xb}(\taub)$ is non-negative definite for each
$\taub$. To check this conditions just note that the Fourier transform of
$\rho_{X_i X_j}(\taub)$, which provides the cross-spectrum between the fields
$X_i(\tb)$ and $X_j(\tb)$, must be a non-negative function.
It is therefore sufficient to check that the $N$-fold Fourier transform of
$\rho_{X_i X_j}(\taub)$ does not take negative values for any pair $i,j$.
In addition, for any fixed $\taub$, $\rhob_{\Xb}(\taub)$ has to be a
non-negative definite matrix and therefore its eigenvalues have to be
non-negative.
This is automatically checked within the field simulation
procedure described in the sections below.

\subsection{Simulating Gaussian multivariate random fields in $R^N$}
\label{sec:gauss}
\label{sec:grf}
To simulate a zero-mean homogeneous multivariate Gaussian field $\Xb(\tb)$
of unit-variance components $X_i(\tb)$ and prescribed cross-correlation
functions, we use the spectral representation that decomposes the field as a superposition of
uncorrelated sinusoids of different frequencies and random amplitudes
\begin{equation} \label{eq:mul1}
X_i(\tb) = \underbrace{\int_{-\infty}^{+\infty} \int_{-\infty}^{+\infty}
\dots \int_{-\infty}^{+\infty}} _{N-fold} {\rm e}^{\iota 2 \pi \kb \cdot \tb}
~d Z_i(\kb), \qquad i=1,2,\ldots,M;
\end{equation}
where $\iota = \sqrt{-1}$, $\kb \cdot \tb$ is the inner product between the
wave number and position vectors, and $d Z_i(\kb)$ are random orthogonal
increments satisfying \citep{pri81}
\begin{eqnarray} \label{eq:mul2}
E[\,dZ_i(\kb)\,] &=& 0, \nonumber\\
E[\,dZ_i(\kb)~dZ^*_j(\kb')\,] &=& 0, ~~~~ \kb \neq \kb', ~~ i,j=1,\ldots,
M, \\ E[\,dZ_i(\kb)~dZ^*_j(\kb)\,] &=& S_{ij}(\kb)~\db\kb, \nonumber
\end{eqnarray}
where ``$~{}^*~$'' indicates the complex conjugate operator, $\db\kb
= dk_1 \cdot dk_2 \cdot \ldots \cdot dk_N$,
and $S_{ij}(\kb)$ is the cross-spectrum of the fields $X_i(\tb)$ and
$X_j(\tb)$.
If the field is Gaussian, then $dZ_i(\kb)$ is an
$M$-dimensional zero-mean complex Gaussian random variable with correlation
matrix determined by the cross-spectrum
\begin{equation}
\Sb_{\Xb}(\kb) = \left(
\begin{array}{ccc}
S_{11}(\kb) & &S_{1 M}(\kb) \\
\vdots & \ddots & \vdots \\
S_{M 1}(\kb) & &S_{M M}(\kb)
\end{array} \right),
\end{equation}
for any fixed $\kb$.
Increments corresponding to different $\kb$'s are independent.

A similar spectral representations for homogeneous random fields defined on
a discrete parameter space is \citep[e.g.,][]{rua98}
\begin{equation} \label{eq:mul3}
X_i(\tb) = \underbrace{\sum \sum \dots \sum}_{N-fold}
{\rm e}^{\iota 2 \pi \kb \cdot \tb} \Delta Z_i(\kb); \qquad i=1,2,\ldots,M,
\end{equation}
(the sums are understood to be over all the discrete wave numbers) where
\begin{eqnarray} \label{eq:mul4}
E[\,\Delta Z_i(\kb)\,] &=& 0, \nonumber\\
E[\,\Delta Z_i(\kb)~ \Delta Z^*_j(\kb')\,] &=& 0, ~~~~ \kb \neq \kb', ~~
i,j=1,\ldots, M, \\
E[\,\Delta Z_i(\kb)~ \Delta Z^*_j(\kb)\,] &=& S_{ij}(\kb)~\Deltab\kb,
\nonumber
\end{eqnarray}
$\Deltab \kb = \Delta k_1 \cdot \Delta k_2 \cdot \ldots \cdot \Delta k_N$.
The discrete representation (\ref{eq:mul3}) approaches the continuous one
(\ref{eq:mul1}) as $|\Deltab \kb|$ approaches zero.
As in the one-dimensional case, the cross-spectrum $\Sb_{\Xb}(\kb)$ is
related to $\rhob_{\Xb}(\taub)$
through the ($N$-fold) discrete Fourier transform. Therefore, under
regularity conditions on $\rhob_{\Xb}(\taub)$,
$\Sb_{\Xb}(\kb)$ is Hermitian and non-negative definite for each value of
$\kb$, and can be factored as
\begin{equation} \label{eq:factor}
\Sb_{\Xb}(\kb) = \Hb(\kb) \Hbs(\kb) = \sum_{l=1}^{M} H_{il}(\kb)
H^*_{jl}(\kb).
\end{equation}
We can also rewrite $\Delta Z_i(\kb)$ in the form
\begin{equation} \label{eq:increment}
\Delta Z_i(\kb) = \sum_{l=1}^{M}  H_{il}(\kb)
~\xi(\kb)~{\rm e}^{\iota \theta_l(\kb)} |\Deltab \kb |^{1/2},
\end{equation}
where $\xi(\kb)$ are independent Rayleigh random deviates and $\theta_l(\kb)$
are independent phase angles uniformly distributed on the interval
$[0, 2\pi]$.
We obtain a random phase representation of the field by inserting this
equation into expression (\ref{eq:mul3})
\begin{equation} \label{eq:mul5}
X_i(\tb) = \underbrace{\sum \sum \dots \sum}_{N-fold} \sum_{l=1}^{M}
{\rm e}^{\iota 2 \pi \kb \cdot \tb} H_{il}(\kb)
~\xi(\kb) ~{\rm e}^{\iota \theta_l(\kb)} |\Deltab \kb |^{1/2}.
\end{equation}
Equation (\ref{eq:mul5}) can be rewritten in the
more revealing form
\begin{equation}\label{eq:IDFT}
X_i(\tb) = {\rm IDFT}_N \left[\, \sum_{l=1}^{M} H_{il}(\kb)
~\xi(\kb) ~{\rm e}^{\iota \theta_l(\kb)}\,
\right] \cdot |\Deltab \kb |^{1/2},
\end{equation}
where ``${\rm IDFT}_N[\cdot]$'' stands for $N$-fold inverse discrete Fourier
transform. This leads to the following procedure for the numerical generation
of a discrete field $\Xb(\tb)$ characterized
by a prescribed $\rhob_{\Xb}(\taub)$:
\begin{enumerate}
\item Determine the cross-spectra $\Sb_{\Xb}(\kb)$ by applying
the $N$-fold discrete Fourier transform to $\rhob_{\Xb}(\taub)$.
\item Factor $\Sb_{\Xb}(\kb)$ as indicated in equation (\ref{eq:factor}).
\item Generate a set of $M$ random Fourier increments from equation
(\ref{eq:increment}) using a set of $M$ independent random phase angles
uniformly distributed over $[0, 2 \pi]$, and a set of $M$ independent
Rayleigh deviates. The phase angles and the Rayleigh deviates at
different wave numbers must be independent.
Rayleigh deviates $\xi(\kb)$ can be generated by the
transformation $\xi(\kb) = \left[-2~\ln u(\kb)\right]^{1/2}$, where $u(\kb)$ is
a uniform random number on the interval $[0, 1]$ \citep{ogo96}.
\item Apply the ($N$-fold) inverse Fourier transform of the random Fourier
increments to obtain a set of $M$ complex (or $2 M$ real) random fields.
\end{enumerate}

Note that the factorization (\ref{eq:factor}) of $\Sb_{\Xb}(\kb)$ is not
unique,  there are a variety of methods
that can be used, some being more stable than others.
One possibility is to obtain
$\Hb(\kb)$ via the singular value decomposition square root of $\Sb_{\Xb}(\kb)$
\begin{equation}
\Hb(k) = \Sb_{\Xb}^{1/2}(k) = \Mb \Jb^{1/2} \Mb^{-1},
\end{equation}
where $\Mb$ is a matrix whose columns are eigenvectors of $\Sb_{\Xb}(k)$, and
$\Jb$ is a diagonal matrix containing the corresponding eigenvalues.
Alternativeley, one can use the Cholesky factorization, which may be more numerically stable \citep{pop98}.
With either factorization it is easy to test if the matrix $\Sb_{\Xb}(\kb)$
is non-negative definite for each $\kb$. Indeed, if
$\Sb_{\Xb}(\kb)$ does not satisfy this condition
the algorithm providing the Cholesky factorization does not converge, and
the matrix $\Jb$ contains one or more negative entries.

Another point to consider is that, although we are interested in the numerical
simulation of real valued random fields,
the suggested procedure does not impose the necessary symmetry restrictions
on the Rayleigh deviates and phase angles in (\ref{eq:mul5}) to avoid complex values
of the fields $X_i(\tb)$.
However, if the Rayleigh deviates and phase angles are
generated with no symmetry restrictions, then the resulting field
$\{ X_i(\tb) \}$ is complex with independent
real and imaginary parts, each with cross-correlation matrix given
by $\rhob_{\Xb}(\taub)/2$ \citep[e.g.,][]{rua98}. Thus, the real and
the imaginary parts of $\sqrt{2}\, X_i(\tb)$
provide two independent real valued realizations of the desired random fields.

\subsection{Simulating Gaussian multivariate random fields on a sphere}

A zero mean random field $R(\rbh)$ on the unit sphere is homogeneous
if the covariance of the field between two different directions,
$\rbh_1$ and $\rbh_2$, depends only on the angular distance that separates
them. In this case the correlation function can be written as
\begin{equation}
\rho_{RR}(\theta) = \frac{{\rm E}
\left[\,R(\rbh_1)\,R(\rbh_2)\,\right]}{\sigma^ 2_{RR}},
\end{equation}
where $\cos \theta = \rbh_1\cdot\rbh_2$, and $\sigma^2_{RR} =
{\rm E}\,[\,R(\rbh_1)^2\,]={\rm E}\,[\,R(\rbh_2)^2\,]$. The field has
the spectral representation
\begin{equation}
R(\rbh) = \sum_{\ell=0}^{\infty}\sum_{m=-\ell}^{\ell}
a_{\ell,m}\,Y_{\ell,m}(\rbh),
\end{equation}
where $\{ Y_{\ell,m} \}$ is a basis of spherical harmonics.
The coefficients $\{\,a_{\ell,m}\,\}$ are zero mean complex uncorrelated
random variables with variance that depends only on $\ell$
\begin{equation}
{\rm E}\left[\,a_{\ell,m}\,a^*_{\ell,m}\,\right] = \sigma_\ell^2,
\end{equation}
and $\sum_\ell \sigma_\ell^2 < \infty$. Similarly, a zero mean multivariate random
field $\Rb(\rbh) = \{ R_1(\rbh),...,R_M(\rbh)\}$ is homogeneous if each $R_i$ is
 a homogeneous field and
the pair-wise cross-correlations depend only on the angular distance. In this case
the cross-correlation
function is as in (\ref{eq:rhos}) but with $\theta$ taking the place of $\tau$.

The spectral representation of the cross-correlation function of $\Rb$ is
\citep{yag86}
\begin{equation}
\rhob_{\Rb}(\theta) = \sum_\ell
\frac{2\ell+1}{4\pi}\,\Sb_\ell\,P_\ell(\cos\theta),
\end{equation}
where $P_\ell$ is the Legendre polynomial of degree $\ell$ and
$\{\,\Sb_\ell\,\}$ are non-negative definite matrices.
The corresponding spectral representation of the multivariate field is
\begin{equation}
\Rb(\rbh) = \sum_{\ell,m} \Ab_{\ell,m}\,Y_{\ell,m}(\rbh),
\end{equation}
where $\Ab_{\ell,m}$ are $M\times 1$ zero mean complex random vectors with
\begin{eqnarray}
{\rm E}\left[\,\Ab_{\ell,m}\,\Ab_{\ell,m}^*\,\right] & = & \Sb_\ell,\\
{\rm E}\left[\,\Ab_{\ell,m}\,\Ab_{\ell,m}^*\,\right] & = &
{\bf 0}\quad\quad\mbox{for $\ell\neq \ell'$ or $m\neq m'$}.
\end{eqnarray}
Each matrix $\Sb_\ell$ can be factored as in equation (\ref{eq:factor}):
\(\Sb_\ell = \Hb_\ell\Hb_\ell^*.\)
As mentioned in the previous section, different factorizations lead to spectral
estimates with different variance characteristics. Any such factorization
leads to representations like (\ref{eq:mul5}) and (\ref{eq:IDFT}) for Gaussian
fields in terms of independent uniform random phases
$\theta_s(\ell,m)$ and independent Rayleigh deviates $\xi_s(\ell,m)$
\begin{eqnarray}
R_i(\rbh) & = & \sum_{\ell,m}\,\Big [\,
\sum_s(\Hb_\ell)_{ks}\,\xi_s(\ell,m)\,e^{\iota\theta_s(\ell,m)}\,\Big
]\,Y_{\ell,m}(\rbh)\\ & = & \mbox{SFT} \Big [\, \sum_s
(\Hb_\ell)_{ks}\,\xi_s(\ell,m)\,e^{\iota\theta_s(\ell,m)}\,\Big ];\qquad
i=1,...,M,
\end{eqnarray}
where SFT stands for spherical Fourier transform. For discrete
implementations of the spherical Fourier transform see, for example,
\cite{dilts85} and \cite{driscoll94}.

\section{Examples}
\label{sec:examples}
\begin{figure}[h]
\plotone{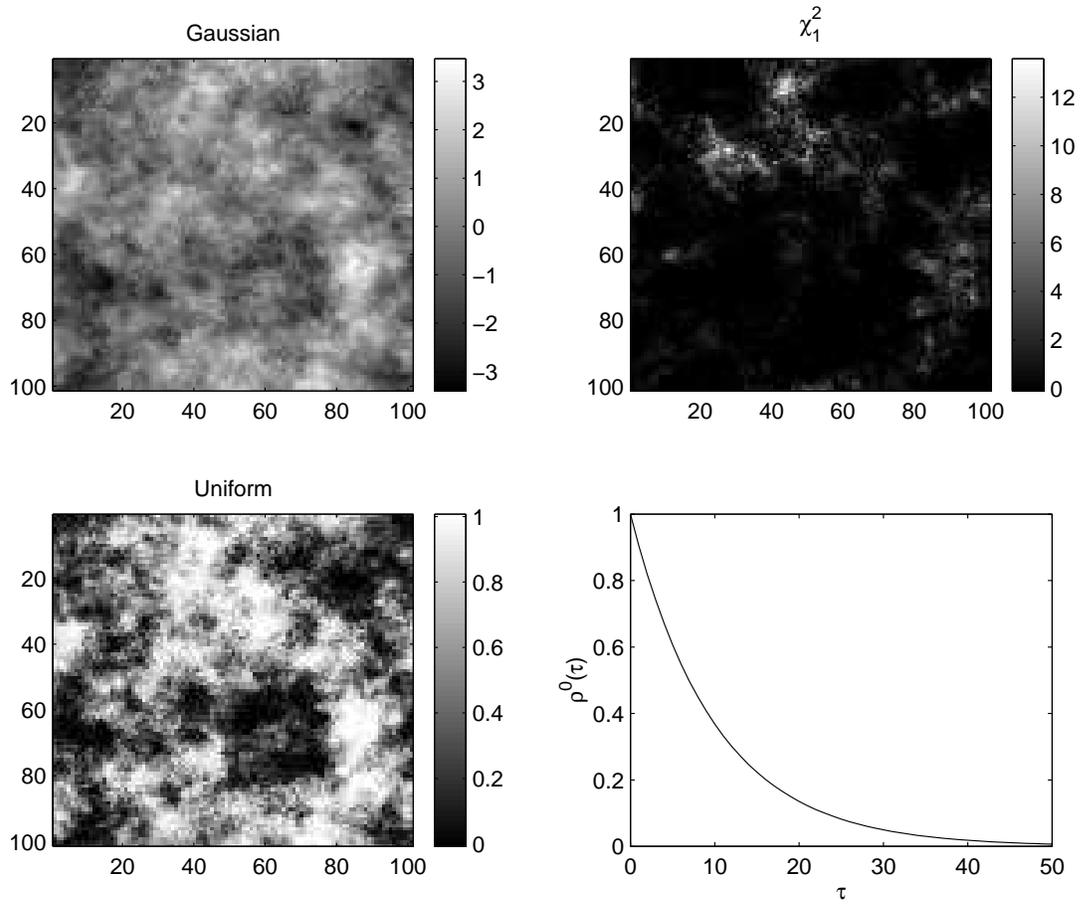}
\caption{Realizations of three correlated random
fields with prescribed marginal distributions and cross-correlation structure
given by equation (\ref{eq:correlation}). The correlation function
$\rho^0(k)$ is also shown for reference.}
\label{fig:cfields}
\end{figure}

We start with a simple numerical simulation.
Figure \ref{fig:cfields} shows a simulation of three two-dimensional,
isotropic,  random fields with a standard Gaussian $N(0,1)$, a $\chi^2_1$,
and a uniform $U(0,1)$ as marginals, respectively. To illustrate different degrees of
correlation in the simulation, we use the cross-correlation matrix
\begin{equation} \label{eq:correlation}
\rhob_{\Rb}(\tau) = \rho^0(\tau) \cdot \left(
\begin{array}{ccc}
1 & 0.3 & 0.9 \\
0.3 & 1 & 0.4 \\
0.9 & 0.4 &1
\end{array}
\right).
\end{equation}

The procedure used is the one described in Section \ref{sec:practical} with
a matrix $\Hb(\kb)$ obtained by a Cholesky factorization of
$\Sb_{\Xb}(\kb)$. The marginal distributions in the three maps are very
different; the Gaussian is symmetric about zero, the $\chi^2$ is
nonnegative and right skewed, the uniform is bounded between zero and one.
However, they all have the same autocorrelation function and there is a
clear correlation between every pair of maps; the correlation is stronger
between the Gaussian and the uniform because of the 0.9 coefficient in
$\rhob_{\Rb}$.

\subsection{Simulating the components of {\it ISO} far-IR maps}

{\it VAW} presented simulations of ISO far-IR maps but their method
was unable to integrate the correlation between the field of localized point
sources and the diffuse background. This integration is needed
since in the observed data the location and the intensity
of the brightest sources are correlated  with the brightest regions in the
background. As a result, the methodology presented in {\it WAV}
is unable to satisfactorily reproduce the localized source features of the
original ISO map on large scale.

\begin{figure}[h]
\plotone{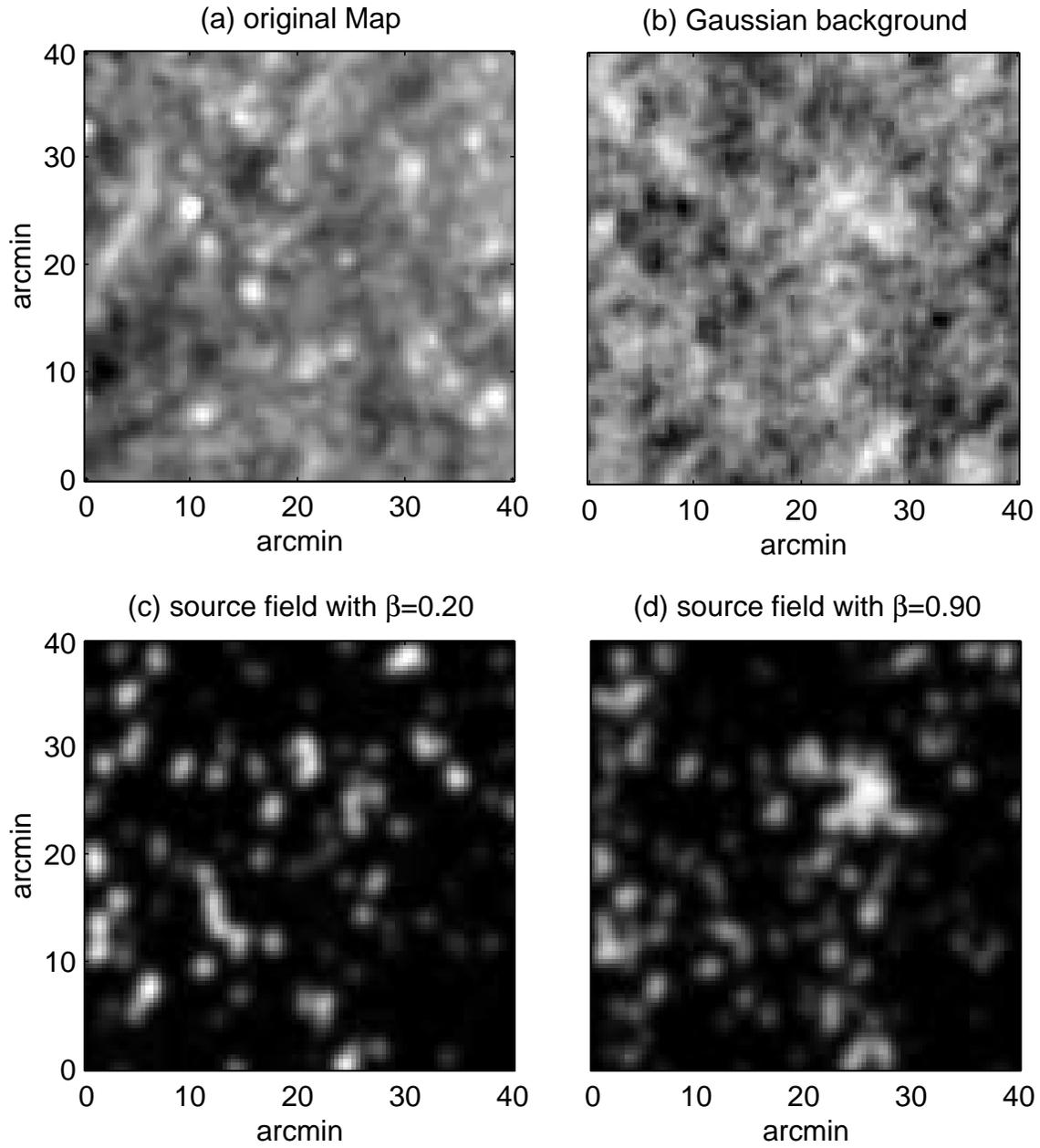}
\caption{Panel (a) shows the original ISO far-IR map.
The simulated Gaussian background map obtained via Fourier phase
randomization is shown in (b).
Panels (c) and (d) show simulated source fields for $\alpha=0.20$ and
$\alpha=0.90$, respectively.}
\label{fig:map} 
\end{figure}

We now show that the methods presented in the previous sections
can account for this type of correlation information and efficiently
exploit it in generating the simulated map.
Figure \ref{fig:map} shows simulations of the random background and source
components of a far-IR map obtained by the following procedure.
Figure \ref{fig:map}a is the {\it observed} map, recorded by the
ISOPHOT camera \citep{lem96} on board the ISO satellite \citep{kes96}
at $175$ $\mu$m, with a projected size of roughly $40' \times 40'$
\citep{dol01}. This map is used to generate a Gaussian background field
(Figure \ref{fig:map}b)
via Fourier phase randomization. If physical models of the diffuse background
were available, they could also be used to simulate this field.

It is well known that extreme values of distributions can be used to generate
point processes. For example, consider $X_1,...,X_n$ uncorrelated Gaussian
random variables. Define $N$ as the number of $X_i$ above a threshold $\tau$ and
$p=P[X_i>\tau]$.
Then $N$ is a Binomial $B(n,p)$ that can be approximated by a Poisson with $\lambda=np$ for $n$ large
and $p$ small.
In our case we need more than a point process for the location of the sources, we also need a distribution
for their intensity. For this reason we have adopted a different approach based on Gaussian mixtures.
These distributions have been successfully used to model outliers and other
types of nonstationary behavior in time series \citep{peel00,lemr96}.
Recall that the distribution of a $k$-component Gaussian mixture is of the form
$F(x) = \sum_{i=1}^k \alpha_i G_i(x)$, where the $\alpha_i\ge 0$ ($\sum_i \alpha_i = 1$) define
the probabilities of sampling from each of the Gaussian distributions $G_i$ (note that a Gaussian
mixture with more than one component is not Gaussian).

For the ISO example we have used a mixture of two Gaussian components for the marginal distribution
of the pixels.
The idea is the following. The location of the extreme values of this mixture define the point
process, the intensities are defined by the Gaussian distribution of the first mixture component.
The second component generates an essentially zero background.
For example, Figure 4 shows a realization of the Gaussian mixture used for Figure 3c (the
intensities in the original ISO map were normalized to unit standard deviation).
We also define a cross-correlation function to correlate the location and intensities
of the sources with the Gaussian background in Figure 3b.

Since we do not have a separate image of the source field,
in order to fix the parameters of the mixture, we identified point sources
by choosing peaks in the original ISO image. The mean and
standard deviation of the selected sources were used to define the Gaussian component
with a fraction $\alpha$ equal to the proportion of pixels identified as sources.
We used a Gaussian centered at zero with a small standard deviation for the second mixture
component (see Figure 4). This means that a proportion $1-\alpha$ of the simulated
observations are essentially zero. The rest of the observations have an intensity spread about the
mean of the observed sources.
In practice the Gaussians are truncated to avoid numerical instabilities in
the integral transform (\ref{eq:map1}).

Finally we have to choose a cross-correlation matrix.
To illustrate the correlation of the sources with the bright
regions in the background, we used the cross-correlation function
\begin{equation}
\rhob(\tau) = \left(
\begin{array}{ccc}
 \rho_0(\tau) & \beta\, \rho_0(\tau)\,\\
\beta\, \rho_0(\tau) & B(\tau)\\
\end{array}
\right),
\end{equation}
where $\rho_0$ is the auto-correlation function of the original ISO map, $\beta\in [-1,1]$
is a constant, and $B(\tau)$ is the correlation function introduced by the instrument's smoothing.
The choice of this form for $\rhob(\tau)$ is based on a lack of sufficient
information about the source field. In particular, the cross-correlation of the background and source fields is assumed
to be a multiple of $\rho_0$ only for illustrative purposes; there are of course applications where
this is not true, see for example \cite{mo96}. For the same reason, we decided to assume that the correlations
in the source field were introduced only by the Gaussian beam smoothing of the instrument.
Figures \ref{fig:map}c,d show fields of sparse localized sources for $\beta=0.2
0$ and $\beta=0.90$, respectively. In both cases, the check described in Section \ref{sec:invprob}
has confirmed that $\rhob(\tau)$ is a valid cross-correlation
matrix for each $\tau$. It is clear that the correlation of the locations and intensities
of the sources with the background map of Figure \ref{fig:map}b increases with $\beta$.

The application of this methodology for the realistic simulation of far-IR maps recorded by the
array cameras on board the {\it Herschel} satellite \citep{pil01} is under investigation.
The combination
of both physical models and
observations is compelling in determining the types of distributions and
cross-correlation functions that are appropriate for the particular
application under investigation.

\section{Summary}
We have described spectral methods for the simulation of homogeneous
non-Gaussian multivariate random fields with prescribed cross-correlation and
marginal distributions. The basic idea is to apply pointwise transformations
to homogeneous Gaussian fields to guarantee the homogeneity of the field.
The transformations are chosen to obtain the desired marginal distributions
and cross-correlation function.
The proposed methodology works for a wide range of cross-correlation functions
and marginal distributions of general interest. Among these, some
Gaussian mixture models may be useful to model fields of localized sources.

We have illustrated the methods by simulating sky maps with
background and source components that are physically linked. The same methods
can also be applied to simulate and/or analyse more complex maps, such
as those from all-sky surveys of the Planck mission
(Mandolesi et al., 1998; Puget et al., 1998).
The objective of the Planck Satellite is to observe the Cosmic Microwave Background
anisotropies which account, however, for only a very small part of the much
more conspicuous backgrounds, either Galactic and Extragalactic.
Sky maps will be produced at nine different frequencies but strong correlations
are expected among different maps because of the wide frequency distribution
of the expected emissions. Many algorithms have been already tested to recover
as much information as possible from the observed maps (see e.g. Maino
et al., 2001 and references therein). But to extract important cosmological
information the methods should provide information on all the contributing components
and should be able to model the correlations among them.
The methods we have presented here are based on theoretically simple assumptions but
do require data, or other available physical information, to select the correct
model parameters (cross-correlation function, marginal distributions, Gaussian mixture
parameters, etc.). Appropriate modeling of more complex sky maps is the goal of
future research.

\clearpage

\begin{figure}[h]
\plotone{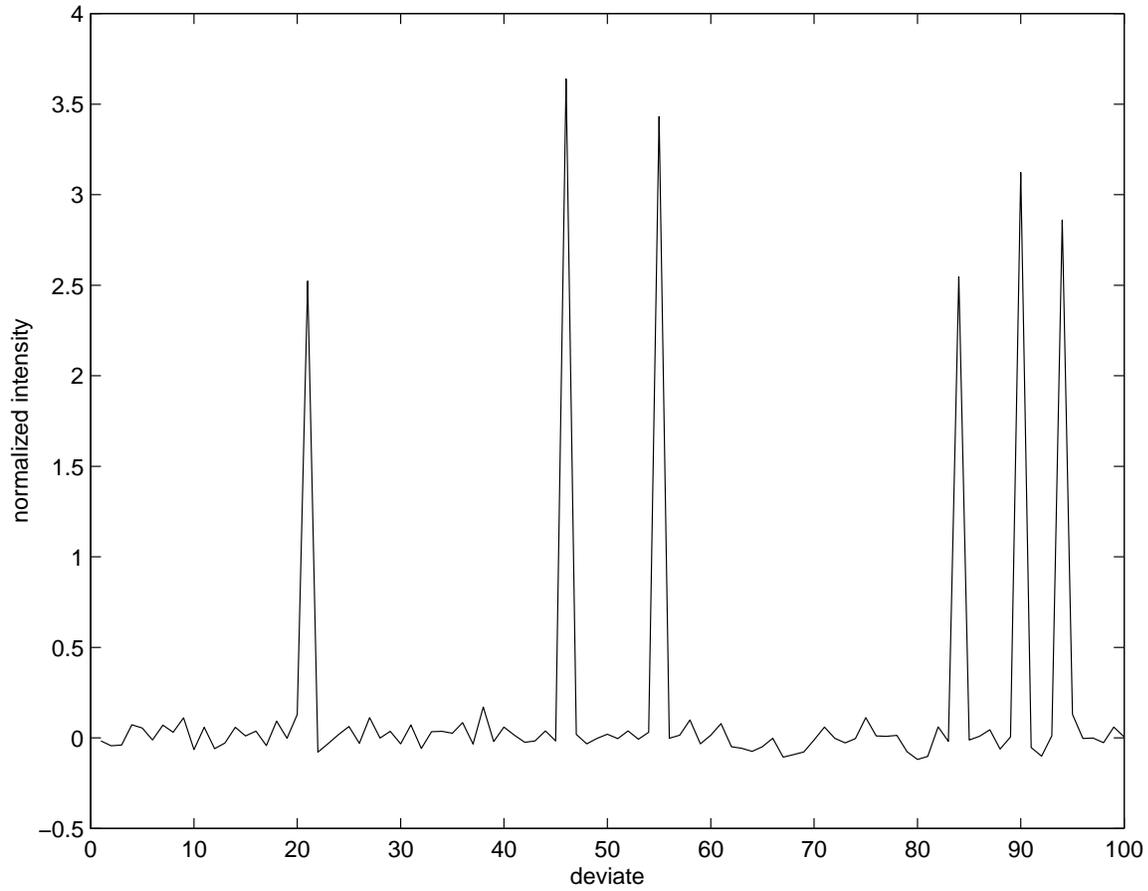}
\caption{One-dimensional realization of a sequence
of uncorrelated deviates from the two-component Gaussian mixture used for Figure 3 (see text).
The intensity was based on the ISO map normalized to unit standard deviation.}
\label{fig:mixture} 
\end{figure}


\begin{thebibliography}{}
\bibitem[Avelino et al. (1998)]{avelino} Avelino, P.P., Shellard, E.P.S., Wu, J.H.P.
\& Allen, B. 1998,
\apj, 507, L101
\bibitem[Diggle (1983)]{dig83} Diggle, P.J 1983, Statistical Analysis of Spatial Point Patterns
(London: Academic Press)
\bibitem[Dilts (1985)]{dilts85} Dilts, G. A. 1985, Journal of
Computational Physics, 57, 439
\bibitem[Dole, et al. (2001)]{dol01} Dole, H., et al. 2001, \aap, 372, 364
\bibitem[Driscoll \& Healy (1994)]{driscoll94} Driscoll, J. R. \& Healy, D.
1994, Adv. in Appl. Math., 15, 202
\bibitem[Kessler et al. (1996)]{kes96} Kessler, M.F., et al. 1996,
\aap, 315, L27
\bibitem[Le, Martin \& Raftery (1996)]{lemr96} Le, N. D., Martin, D. \& Raftery, A. E. 1996,
Journal of the American Statistical Association, 91, 1504
\bibitem[Lemke, Klaas \& Abolins (1996)]{lem96} Lemke, D, Klaas, U. \& Abolins,
J. 1996, \aap, 316, L64
\bibitem[Maino et al., (2001)]{mai01} Maino D., et al. 2001, \mnras, {\it submitted} (astro-ph/0108362)
\bibitem[Mandolesi et al., (1998)]{man98} Mandolesi N., et al. 1998, Planck
Low Frequency Instrument, a proposal submitted to ESA
\bibitem[Martinez \& Saar (2001)]{martinez} Martinez, V.J. \& Saar, E. 2001,
Statistics of the Galaxy Distribution (New York: Chapman \& Hall)
\bibitem[McLachlan \& Peel (2000)]{peel00} McLachlan, G. \& Peel, D.A.
2000, Finite Mixture Models (New York: Wiley)
\bibitem[Mo \& White (1996)]{mo96} Mo, H.J. \& White, S.D.M. 1996, MNRAS, 282, 347
\bibitem[Ogorodnikov \& Prigarin (1996)]{ogo96} Ogorodnikov, V.A. \&
Prigarin, S.M. 1996, Numerical Modelling of Random Processes and Fields:
Algorithms and Applications (Utrecht: VSP)
\bibitem[Peebles (1999)]{peebles} Peebles, P.J.E. 1999, \apj, 510, 531
\bibitem[Pilbratt (2001)]{pil01} Pilbratt, G. 2001, The FIRST ESA Cornerstone
Mission, in IAU Symp. 204, The Extragalactic Infrared Background and its
Cosmological Implications, August 2000 Manchester, eds. M. Harwit
\& M.G. Hauser (San Francisco: ASP), 39
\bibitem[Popescu, Deodatis \& Prevost (1998)]{pop98} Popescu, R., Deodatis, G.
\& Prevost, H. 1998, Prob. Engng. Mech., 13, 1
\bibitem[Priestley (1981)]{pri81} Priestley, M.B. 1981, Spectral Analysis and
Time Series (London: Academic Press)
\bibitem[Puget et al., (1998)]{pug98} Puget, J.L., et al. 1998,
High Frequency Instrument for the Planck Mission, a proposal submitted to ESA
\bibitem[Ruan \& McLaughlin (1998)]{rua98} Ruan, F. \& McLaughlin, D. 1998,
Advances in Water Resources, 21, 385
\bibitem[Vio, Andreani \& Wamsteker (2001)]{vio01} Vio, R., Andreani, P.
\& Wamsteker, W. 2001, \pasp,  113, 1009 (VAW)
\bibitem[Yaglom (1986)]{yag86} Yaglom, A. M. 1986, Correlation Theory
of Stationary Random Functions I (New York: Springer-Verlag)

\end{thebibliography}
\end{document}